\documentclass[11pt,epsf]{article}

\usepackage{epsfig}
\usepackage{amssymb}
\usepackage{graphicx}
\usepackage{color}
\usepackage{subfigure}

\begin{document}

\baselineskip 6mm
\renewcommand{\thefootnote}{\fnsymbol{footnote}}

%------------ Chanyong Park's macro, etc  -----------

\newcommand{\nc}{\newcommand}
\newcommand{\rnc}{\renewcommand}

%\headheight=0truein
%\headsep=0truein
%\topmargin=0truein
%\oddsidemargin=0truein
%\evensidemargin=0truein
%\textheight=9.5truein
%\textwidth=6.5truein

%\rnc{\baselinestretch}{1.24}    % 1.5 spacing btwn text lines
%\setlength{\jot}{6pt}       % spacing btwn the rows of an eqnarray
%\rnc{\arraystretch}{1.24}   % spacing btwn the rows of a non-eqn array

%%%%%%%%%%%%%%%%%%%%%% Equation Numbering %%%%%%%%%%%%%%%%%%%%%%%
%\makeatletter \rnc{\theequation}{\thesection.\arabic{equation}}
%\@addtoreset{equation}{section} \makeatother

%%%%%%%%%%%%%%%%%%%%%%%%%%%%%%%%%%%%%%%%%%%%%%%%%%%%%%%%%%%%%%%%%
%                                                               %
%                NEW COMMANDS AND MACROS                        %
%                                                               %
%%%%%%%%%%%%%%%%%%%%%%%%%%%%%%%%%%%%%%%%%%%%%%%%%%%%%%%%%%%%%%%%%

\newcommand{\tcb}{\textcolor{blue}}
\newcommand{\tcr}{\textcolor{red}}
\newcommand{\tcg}{\textcolor{green}}

%%%%% Simplify some frequently used LaTeX commands %%%%%

\def\be{\begin{equation}}
\def\ee{\end{equation}}
\def\ba{\begin{array}}
\def\ea{\end{array}}
\def\bea{\begin{eqnarray}}
\def\eea{\end{eqnarray}}
\def\nn{\nonumber\\}

%%%%%  Temporary notation %%%%

\def\ct{\cite}
\def\la{\label}
\def\eq#1{Eq. (\ref{#1})}

%%% Greek letters %%%

\def\a{\alpha}
\def\b{\beta}
\def\g{\gamma}
\def\G{\Gamma}
\def\d{\delta}
\def\D{\Delta}
\def\ep{\epsilon}
\def\et{\eta}
\def\ph{\phi}
\def\Ph{\Phi}
\def\ps{\psi}
\def\Ps{\Psi}
\def\k{\kappa}
\def\l{\lambda}
\def\L{\Lambda}
\def\m{\mu}
\def\n{\nu}
\def\th{\theta}
\def\Th{\Theta}
\def\r{\rho}
\def\s{\sigma}
\def\S{\Sigma}
\def\ta{\tau}
\def\o{\omega}
\def\O{\Omega}
\def\pr{\prime}

%%%%% Mathematical Symbols

\def\half{\frac{1}{2}}

\def\goto{\rightarrow}

\def\na{\nabla}
\def\grad{\nabla}
\def\curl{\nabla\times}
\def\div{\nabla\cdot}
\def\pa{\partial}

\def\bra{\left\langle}
\def\ket{\right\rangle}
\def\lb{\left[}
\def\lc{\left\{}
\def\ls{\left(}
\def\ln{\left.}
\def\rn{\right.}
\def\rb{\right]}
\def\rc{\right\}}
\def\rs{\right)}

\def\vac#1{\mid #1 \rangle}

%%%%  Special symbol

\def\td#1{\tilde{#1}}
\def\check{ \maltese {\bf Check!}}

%%%%% Roman pont in math

\def\Tr{{\rm Tr}\,}
\def\det{{\rm det}}

%%%%% Special format

\def\bc#1{\nnindent {\bf $\bullet$ #1} \\ }
\def\ch {$<Check!>$ }
\def\ss {\vspace{1.5cm}}

\begin{titlepage}
%---------------- preprint number ---------------
\hfill\parbox{5cm} { }

\vspace{25mm}

\begin{center}
%------------------------ title ------------------------
{\Large \bf Meson spectra in a gluon condensate background}
%---------------- authors and addresses ----------------
\vskip 1. cm
  {
  Yumi Ko$^a$\footnote{e-mail : koyumi@sogang.ac.kr},
  Bum-Hoon Lee$^{ab}$\footnote{e-mail : bhl@sogang.ac.kr},
  and Chanyong Park$^a$\footnote{e-mail : cyong21@sogang.ac.kr} }

\vskip 0.5cm

{\it

$^a\,$ Center for Quantum Spacetime (CQUeST), Sogang University, Seoul 121-742, Korea \\
$^b\,$ Department of Physics, Sogang University, Seoul 121-742, Korea }\\

\end{center}

\thispagestyle{empty}

\vskip2cm

%----------------------- abstract ----------------------

\centerline{\bf ABSTRACT} \vskip 4mm
We study the spectra of light mesons as well as the dissociation
of a quarkonium and monopole-anti monopole bound state in a gluon condensate background.
In order to describe the confining phase we introduce an IR cutoff in two ways,
namely the hard wall and the braneless approaches. We find that the meson spectra strongly
depend on the gluon condensate in the braneless approach, while
they do not in the hard wall model. The latter result appears to be a more probable feature of QCD.

\vspace{1cm}

\vspace{2cm}
%PACS numbers:

%\today

\end{titlepage}

\renewcommand{\thefootnote}{\arabic{footnote}}
\setcounter{footnote}{0}

\tableofcontents
%%%%%%%%%%%%%%%%%%%%%%%%%%%%%%%%%%%%%%%%%%%%%%%%%%%%%%%%%%%%%%%%%%%%%%%%%%%%%%
%                                                                            %
%   Sec.  Introduction                                                       %
%                                                                            %
%%%%%%%%%%%%%%%%%%%%%%%%%%%%%%%%%%%%%%%%%%%%%%%%%%%%%%%%%%%%%%%%%%%%%%%%%%%%%%

\section{Introduction}

Low energy QCD is an important and challenging field to study in spite of the difficulties
in dealing with its nonperturbative nature. Recently, there have been many attempts, termed holographic QCD, to
understand such strongly interacting systems based on the AdS/CFT correspondence \ct{Maldacena:1997re,Policastro:2002se,Herzog:2006ra,Sakai:2004cn,{Erlich:2005qh},
Karch:2006pv,{Da Rold:2005zs},Sin:2009dk,Gursoy:2008za,{Babington:2003vm},{Evans:2004ia},{Kruczenski:2003uq}}. The holographic idea was applied to construct a dual gravity
model for QCD incorporating chiral symmetry breaking in Refs.
\cite{{Erlich:2005qh},Karch:2006pv,{Da Rold:2005zs},{Evans:2004ia},{Kruczenski:2003uq}}. By introducing an IR cutoff or a dilaton potential
in the AdS space to describe the quark confinement, the masses of the various mesons and decay constants
were investigated in these works.

In the confining phase there exist important vacuum quantities such as
the baryon number density,
the chiral condensate and gluon condensate which can affect these physical properties such as the meson spectrum.
Therefore it is  of interest to properly implement these effects in
the holographic QCD model. To do so, we should find the dual geometry in the confining phase, include these
vacuum effects and investigate how they change the physical quantities such as the
mass spectrum, the deconfinement temperature, and other relevant quantities.

For the case of a finite baryon density, the thermally charged AdS space was proposed as
the dual geometry of the hadronic phase with a finite baryonic chemical potential or
number density \cite{Lee:2009by}. In this background, the light meson and
quarkonium spectra and the deconfinement phase transition were studied in Refs.
\cite{Park:2009nb,Jo:2009xr,BoschiFilho:2002ta}.
The chiral condensation is dual to a massive scalar field in the bulk. In Ref. \ct{Kim:2007em}
the first correction of the massive scalar field to the Hawking-Page transition temperature was
investigated in a  background without backreaction of the massive scalar field.
As the dual geometry containing the chiral condensate
is still not known, it would be an interesting problem in itself to find the full solution including
the backreaction of the scalar field. In the case with a specific scalar potential
\cite{{Shock:2006gt}}, meson spectra has been studied in a background containing a backreaction
of the massive scalar field.
On the other hand, the dual geometry including the gluon condensate can be obtained
by analytically solving a gravity theory coupled to a dilaton, i.e. a massless scalar field
\ct{Gubser:1999pk,Kim:2007qk,Kehagias:1999tr,Csaki:2006ji,Evans:2006dj}.
Usually, there are two types of solutions, the dilaton
wall solution and the dilaton black hole solution. These correspond to
the confining and deconfining backgrounds, respectively. In the deconfining background
the effect of the gluon condensate on the quarkonium spectrum as well as the deconfinement
phase transition  have both been already investigated in Ref. \cite{Kim:2008ax}.

In this work, we study the effects of gluon condensation on the light meson spectra and
the dissociation of the quarkonium and monopole bound state in a confining background. As we will
see, the dilaton wall solution becomes divergent at $z=z_c$. However, it
is known as a ``good singularity'' which does not appear in the action because the divergent
terms coming from the curvature scalar and the dilaton kinetic term are exactly
canceling each other. In order to calculate the mass spectrum of mesons in a confining
background, we need to take into account an IR cutoff. For this, we can either take
the singular point $z_c$ as an intrinsic IR cutoff or insert by hand a hard wall at $z=z_{IR}$ in
front of $z_c$. We refer to the first procedure as the ``braneless'' approach, and the second
as the ``hard wall'' approach.

In QCD, the effect of the baryon density and the chiral condensate on physical properties
of quarks are expected to be stronger than the gluon condensation effect. However, the gluon condensate background is of theoretical interest because it may be solved analytically.
Moreover, this work may shed further light on studies of the chiral condensate dependence of
the meson spectra in holographic QCD, and on investigations of superconductivity in AdS/CMT.
Thus it is  meaningful to study the behavior of physical
quantities depending on the gluon condensate. To describe more realistic phenomena,
the model should be improved by incorporating the effect of chiral condensates and the
baryon number density in the geometry and should be studied elsewhere.

We organize our paper as follows: in Sec. 2, we provide a review of  gluon condensation
and its dual geometry. Next we consider the mass spectrum of various types of mesons, in the presence (Sec. 3) and absence (Sec 4.) of a hard wall as an IR cutoff. We will call these
two approaches the hard wall approach and the braneless approach, respectively. Finally, we
discuss our results in  sec. 5.

%%%%%%%%%%%%%%%%%%%%%%%%%%%%%%%%%%%%%%%%%%%%%%%%%%%%%%%%%%%%%%%%%%%%%%%%%%%%%%
%                                                                            %
%   Sec.  Contents                                                           %
%                                                                            %
%%%%%%%%%%%%%%%%%%%%%%%%%%%%%%%%%%%%%%%%%%%%%%%%%%%%%%%%%%%%%%%%%%%%%%%%%%%%%%

\section{Gluon condensate background}

In this section, we will shortly review gluon condensation and its dual geometry
\ct{Gubser:1999pk,Kim:2007qk,Kehagias:1999tr,Csaki:2006ji}. %explaning our conventions.
In holography, the gluon condensation is dual to the dilaton field on the gravity side.  Thus
we consider a five-dimensional Euclidean gravity action with a cosmological constant, $\L = - \frac{6}{R^2} $, and  a dilaton field \ct{Csaki:2006ji,Kim:2007qk}
\be \la{act:background}
S = \frac{1}{2 \k^2}  \int d^5 x \sqrt{G}  \ls  - {\cal R} + 2 \L
+ \frac{1}{2} \pa_{M} \ph \pa^M \ph \rs .
\ee
Here $\k^2$ is the five-dimensional Newton constant and $R$ is a curvature radius.
The Einstein equation and the equation of motion for the scalar field
read
\bea    \la{eq:einstein}
{\cal R}_{MN} - \half G_{MN} {\cal R} + G_{MN} \L &=& \frac{\k^2}{2} \lb
\pa_M \ph \pa_N \ph - \half G_{MN} \pa_P \ph \pa^P \ph \rb , \nn
0 &=& \frac{1}{\sqrt{G}} \pa_M \sqrt{G} G^{MN} \pa_N \ph .
\eea

As shown in Ref. \ct{Kim:2008ax}, there exist two solutions. One is a dilaton wall solution, i.e. a deformation of AdS spacetime. It corresponds to the confining phase with  gluon condensation.
The other is a dilaton black hole solution, i.e.  a deformation of the Schwarzschild-type
AdS black hole with a naked singularity at the horizon. The naked singularity gives rise to some problems, e.g. a proper definition of the Hawking temperature. As the aim of this paper is to
study the meson spectra depending on the gluon condensation in the confining phase, we will only
focus on the dilaton wall solution which forms as follows:
\bea    \la{dilatonmetric}
ds^2 &=& \frac{R^2}{z^2} \ls \sqrt{1-c^2 z^8} \d_{\m\n} dx^{\m} dx^{\n} +dz ^2 \rs , \nn
\phi(z) &=& \phi_0 + \sqrt{\frac{3}{2}}\log \left( \frac{1+cz^4}{1-cz^4} \right) ,
\eea
where $\ph_0$ and $c$ are integration constants.

Besides the gluon condensation, there are other important effects in the confining phase such as the chiral condensation
and the quark density effects. In both cases, the number of degrees of freedom is proportional to $N_c N_f$ while  $N_c^2$ for the gluon condensation. Thus, in the limit $N_c \gg N_f$ in confining phase, the effect of gluon condensation on the physical properties  becomes dominant. This regime is what we are interested in this paper.  It would  also be interesting to investigate the effect of chiral condensation for $N_c \sim N_f$, and  will be left for future  work.

 In order to describe the confining behavior of the dual gauge theory, we need to introduce an IR cutoff.
For this purpose, we choose two possible ways, namely the ``hard wall approach'' and the ``braneless approach.''

In the hard wall approach, we insert a hard wall as an IR cutoff in front of the singular point $z_c \equiv \frac{1}{c^{1/4}}$ by hand. In this case, we do not need to worry about the singularity as it is screened by the IR cutoff.
 On the other hand, the singular point $z_c$ itself can be regarded as an IR cutoff, which is the principle of the braneless approach. Its merit is that we do not need to introduce an IR cutoff by hand.
 At a glance, one may wonder whether the logarithmic divergence of the scalar field in \eq{dilatonmetric} at $z_c$
causes some problems for the calculation of the meson spectra. However, as will be shown later, the logarithmic divergence
 is removed as the metric component vanishes at $z_c$.

Near the boundary when $z \rightarrow 0$, the perturbative expansion of the dilaton field
becomes
\be
\ph = \ph_0 + \sqrt{6} \ \frac{z^4}{z_c^4} + {\cal O} (z^8) \,.
\ee
According to the AdS/CFT correspondence, the constant term, $\phi_0$, corresponds to a source
for the gluon condensation operator $\Tr G^2 $ and the coefficient of the normalizable mode
gives the gluon condensate which is\ct{Csaki:2006ji}
\be \la{gluoncond}
\bra \Tr G^2 \ket = \frac{8 \sqrt{3 (N_c^2-1)}}{\pi}  \frac{1}{z_c^4} ,
\ee
where $\frac{1}{\k^2} = \frac{4 (N_c^2 -1 )}{\pi^2 R^3}$ is used. To consider the meson spectra
on the gluon condensate background, we should turn on several bulk field fluctuations on the dilaton wall background. The action of these fluctuations
is given by \cite{Erlich:2005qh,Da Rold:2005zs}
\be \la{act:fluctuation}
\D S = \int d^5 x \sqrt{G} \ \Tr \lb %\frac{1}{g^2} \ls
|D X|^2 - \frac{3}{R^2} |X|^2 % \rs
+ \frac{1}{4 g_5^2} \ls F_L^2 + F_R^2 \rs \ \rb ,
\ee
with
\be
%\frac{1}{g^2} = \a \frac{N_c}{R} \ \ {\rm and} \ \
\frac{1}{g_5^2} = \frac{N_c}{12 \pi^2 R},
\ee
where $R$ and $N_c$ are needed for the correct dimensionality and proper $1/N_c$
suppression, respectively.
Note that
$D$ is a covariant derivative, $D_M X = \pa_M X - i A^{(L)}_M  X  + i X A^{(R)}_M $
and $F^{(L,R)}_{MN} = \pa_M A^{(L,R)}_N - \pa_N A^{(L,R)}_M
- i \lb A^{(L,R)}_M, A^{(L,R)}_N \rb$, where the combinations of $A^L_M$ and $A^R_M$
\bea
V_M &=& \half \ls A^L_M + A^R_M \rs , \nn
A_M &=& \half \ls A^L_M - A^R_M \rs ,
\eea
correspond to the vector meson or axial vector meson, respectively.
The complex scalar field $X$ transforms in the bifundamental representation under
the flavor symmetry group, $SU(N_f)_L \times SU(N_f)_R$, which can be parameterized as
\be
X = \frac{v}{2} e^{2 i \pi} ,
\ee
where the background field $v$ and fluctuation $\pi$ corresponds to the quark mass (or
chiral condensate) and pseudoscalar meson.

\section{Hard wall approach}

In the gluon condensate background as shown in the previous section, there exists a singular
point, which can be considered as an IR cutoff. We will call this set-up a braneless approach
which will be investigated in the next section. In this section, we will consider a
hard wall approach, in which we insert a new IR cutoff by hand in front of the geometric
singular point.

There are several motivations why we introduce the new IR cutoff. The first one
is related to the generalization of this model. For the massless scalar case,
the on-shell gravity action is finite since the divergences from the curvature
and the kinetic term of the scalar field cancel each other exactly. So the free energy of dual
QCD corresponding to on-shell bulk gravity action is finite.
If we add a mass or potential term for the scalar field, the on-shell
gravity action diverges if there exists a geometrical cutoff.
Therefore, the dual free energy is not well defined. This fact
implies that the geometrical cutoff can not be generally considered as
a proper IR cutoff. One
way to cure this problem is to introduce another IR cutoff screening the geometrical
singular point. The second one is how to impose the IR boundary condition at the
geometrical cutoff where the scalar field and curvature diverge. Actually,
the IR boundary condition
plays an important role to determine the relation between various parameters.
If we insert the new IR cutoff, we can easily impose the IR boundary condition although
it still remains to find a physical IR boundary condition explaining
the parameter relations in QCD.

In this paper, we will investigate the light meson and quarkonium spectra in these two
approaches.
For comparing our results with those in Ref. \ct{Erlich:2005qh}, we will start with the
hard wall approach, in which
we insert an IR brane (or hard wall) in front of the singular point $z_c$.
If the position of the hard wall is $z=z_{IR} < z_c$, the range of the $z$-coordinate in the bulk
space is limited to $0 \le z \le z_{IR}$. In this case the singular point lies outside the bulk,
and we do not need to worry about the singularity. In this paper we choose
$z_{IR} =(323 {\rm MeV})^{-1}$ because
it generates the correct $\r$-meson mass in the original hard wall model \ct{Erlich:2005qh}.

\subsection{Vector meson}

Following the AdS/CFT correspondence, the bulk gauge field fluctuations correspond to the
meson in holographic QCD. In this section, we will investigate the vector meson
spectrum by turning on the vector gauge field, which does not mix with
the scalar and axial gauge field. In the axial gauge $V_z=0$, the equations of motion
for the vector gauge field $V_i$ ($i=1,2,3$) become
\be
0 = \frac{1}{\sqrt{G}} \pa_M \sqrt{G} G^{MP} G^{ij} \pa_P V_i  ,
\ee
where $M$ and $P$ are five-dimensional indices.
Using the ansatz,
\be
V_{i} = \int \frac{d^4 k}{(2\pi)^4} \ e^{- i \o_n t + i \vec{p}_n \vec{x}} \  V_i^{(n)} (z) ,
\ee
where $n$ implies the $n$-th excitation mode, the above equations are reduced to
\be \la{eq:vector}
0 = \pa_z^2 V_i^{(n)} - \frac{1 + 3 c^2 z^8}{z(1-c^2 z^8)} \pa_z V_i^{(n)}
+ \frac{m_n^2}{\sqrt{1-c^2 z^8}} V_i^{(n)} \, .
\ee
Here, $m_n^2 = w_n^2 - \vec{p}_n^2$ is a mass squared of the $n$-th excited meson state.
Especially, the first excited state is called $\r$-meson, whose mass is denoted by $m_{\r}$.
To obtain the vector meson mass, we impose the Dirichlet boundary
condition ($V_i^{(n)} = 0$) at the UV fixed point ($z=0$) and the Neumann boundary condition
($\pa_z V_i^{(n)} = 0$) at the IR cutoff ($z=z_{IR}$). Since the range of $z$ is finite, the boundary conditions are satisfied only at some discrete values of $m_{\r}$ for $n=1$.
We find the values of $m_\rho$ depending on gluon condensation numerically, as listed in Table 1. We see that as the gluon condensation increases, the vector meson mass  decreases.

%%%%%%%%%%%%%%%%%%%%%%%%%%%%%%%%%%%%%%%%%%%%%%%%%%
%%%%%%%%%%%%%%%%%%%%%%%%%%%%%%%%%%%%%%%%%%%%%%%%%%
\begin{center}
\begin{tabular}{|c||c|c|c|c|c|c|}
\hline
 $z_c $ {\scriptsize (1/GeV)} & $ \langle {\rm Tr} G^2\rangle$  {\scriptsize (GeV$^4$)}& $ m_\rho$  {\scriptsize (GeV) }& $m_A$  {\scriptsize (GeV)} & $m_\pi$  {\scriptsize (GeV)}\\
\hline \hline
$\infty$ & 0 & 0.7767 & 1.3582 & 0.13961\\
\hline
 1/0.176 &  0.012 & 0.7767 & 1.3583 & 0.13961\\
\hline
 1/0.200 & 0.020 & 0.7767 & 1.3584 & 0.13961\\
\hline
1/0.250 &  0.049 & 0.7762 & 1.3589 & 0.13964\\
\hline
 1/0.280 & 0.077 & 0.7755 & 1.3599 & 0.13970\\
\hline
 1/0.320 & 0.131 & 0.7724 & 1.3612 & 0.13999\\
\hline
\end{tabular}  \\
\vspace{0.3cm}
\noindent Table 1 : Various meson masses depending on the gluon condensation\\
in the hard wall approach. \\
\end{center}
%%%%%%%%%%%%%%%%%%%%%%%%%%%%%%%%%%%%%%%%%%%%%%%%%%

\subsection{Chiral condensate}

The chiral condensation is obtained by solving for $v(z)$, i.e. the expectation value of the complex scalar field $X$.
From the action of \eq{act:fluctuation}, the equation of motion for $v(z)$ is given by
\bea    \la{eqv}
0 &=& \pa_z \lb \sqrt{G} g^{zz} \pa_z v  \rb + \frac{3}{R^2} \sqrt{G} v \nn
%&=& \pa_z \frac{(1-c^2 z^8)}{z^3} \pa_z v + \frac{3 (1-c^2 z^8)}{z^5} v \nn
&=& \pa_z^2 v - \frac{3+5 c^2 z^8}{z(1-c^2 z^8)}\, \pa_z v + \frac{3}{z^2} v\,.
\eea
Its exact solution is
\be \la{esol}
v = A  z \ _2F_1 \lb \frac{1}{4}, \frac{1}{2}, \frac{3}{4},c^2 z^8 \rb +
    B  z^3 \ _2F_1 \lb \frac{1}{2}, \frac{3}{4}, \frac{5}{4}, c^2 z^8 \rb ,
\ee
where $A$ and $B$ are integration constants.
Near the UV cutoff ($z \to 0$), $v(z)$ perturbatively expands as
\be \la{asyv}
v = A  z + B  z^3 + {\cal O} (z^9) ,
\ee
and accords with the solution on the pure AdS background.  Now we can identify the coefficient
of the non-normalizable mode to the current quark mass $m_q$  as
\be
A = m_q ,
\ee
where $R$ is set to one for simplicity, and  the coefficient of
the normalizable mode is related to the chiral condensate as
\be
B= \s \equiv \bra \bar{q} q \ket .
\ee

Notice that each term on the right hand side of \eq{esol} has a logarithmic divergence
at $z=z_c$. As we expand each term near $z_c$, we encounter the divergence which reads
\bea    \la{divergence}
z \ _2F_1 \lb \frac{1}{4}, \frac{1}{2}, \frac{3}{4},\frac{z^8}{z_c^8} \rb &=&
 - 0.19069 z_c \ls \log \frac{\ep}{z_c} - 2.95709 \rs + {\cal O} (\ep),  \nn
z^3 \ _2F_1 \lb \frac{1}{2}, \frac{3}{4}, \frac{5}{4},\frac{z^8}{z_c^8} \rb &=&
- 0.41731 z_c^3 \ls \log \frac{\ep}{z_c} + 0.184502 \rs + {\cal O} (\ep) ,
\eea
where $\ep = z-z_c$. Even though the value of $v$ becomes singular at $z=z_c$, the logarithmic
divergence does not affect our calculation  as the range of $z$
 is restricted to $0 \le z \le z_{IR}$ with $z_{IR} < z_c$.

\subsection{Axial vector meson and pion}
Let us make the following ansatz for the axial gauge field,
\bea
A_{\m} &=& \int d^4 k \ e^{i q x} \ls \bar{A}_{\m} + \pa_{\m} \varphi \rs , \nn
\pi &=& \int d^4 k \ e^{i q x} \pi_q ,
\eea
where $\bar{A}$ means a transverse gauge field in the bulk
satisfying $\pa_{\m} A^{\m} = 0$ and ${\varphi}$ a gauge transformation. The equations of motion for
the axial gauge field and the pseudoscalar fields are
\bea    \la{eqs}
0 &=& \pa_z \ls \frac{\sqrt{1-z^8/z_c^8}}{z} \pa_z \bar{A}_{\m} \rs + \frac{m_A^2}{z} \bar{A}_{\m}
      - \frac{g_5^2 \G(z)}{z^3}  \bar{A}_{\m} ,\nn
0 &=& \pa_z \ls \frac{\sqrt{1-z^8/z_c^8}}{z} \pa_z \varphi \rs + \frac{g_5^2
      \G(z)}{z^3}  \ls \pi_q - \varphi \rs, \nn
0 &=& - m_{\pi}^2 \pa_z \varphi + \frac{g_5^2 \G(z)}{z^2}  \pa_z \pi_q ,
\eea
where $\G(z) = v^2 \sqrt{1-z^8/z_c^8}$ and we set $R=1$.
Since the first equation for the axial vector meson is decoupled from the two other ones, we can numerically solve it and determine the mass of the axial vector meson.
We again impose a Dirichlet boundary condition ($\bar{A}_{\m} = 0$) at the UV fixed point, $z=0$,  and a Neumann boundary condition ($\pa_z \bar{A}_{\m}= 0$) at the IR cutoff, $z=z_{IR}$.
In Table 1, we show the masses of the first excited axial vector meson, varying values
of the gluon condensation. Here, we used $m_q=2.29$MeV and $\s = (327$MeV$)^3$ obtained in Ref. \cite{Erlich:2005qh} to compare our results with the ones of the original hard wall model.

The pion mass can be obtained by solving the last two coupled equations in \eq{eqs}.
We choose three boundary conditions, $\varphi = 0$ and $\pi_q = 0$ at the boundary, and
 $\pa_z \varphi = 0$ at the IR cutoff, $z=z_{IR}$. See also Table 1 for the mass of the pion and mass of the first excited pseudoscalar meson as functions of the gluon condensate.

We find that  the gluon condensate dependence of the various meson masses is rather  weak. However, from a  qualitative point of view,
increasing the gluon condensate leads to an increase of the axial vector meson mass and the pion mass and to a decrease of the vector meson mass.

\subsection{Dissociation of a quarkonium and a monopole bound state}

So far, we have studied mass spectra for various light mesons depending
on the gluon condensate. Let us now consider the quarkonium and the monopole-anti monopole bound state, described by a fundamental open string and
a D1-string respectively. In this section, we investigate the gluon condensate  dependence of their dissociation length in the confining phase.

First, we consider the quarkonium composed of two heavy quarks.
The fundamental string action in the gluon condensated
background \cite{Gubser:1999pk,Csaki:2006ji,Kim:2007qk} is given by
\be
S_{F1} = \frac{1}{2 \pi \a'} \int d^2 \s \ e^{\ph / 2} \sqrt{\det\, h_{\a \b}} ,
\ee
where $h_{\a \b}$ is the induced metric on the string worldsheet.
For simplicity, we take
\be \la{simple}
\frac{R^2}{2 \pi \a'}=1 \quad {\rm and} \quad \ph_0=0 .
\ee
At first we consider the string configuration connecting two heavy quarks
with the following ansatz
\be
\ta=t, \quad \sigma=x^1 \equiv x  \quad {\rm and} ~~ z=z(x) ,
\ee
the fundamental string action in the background of \eq{dilatonmetric} is then reduced to
\be
S_{F1} = \int_{-T/2}^{T/2} dt \int_{-r/2}^{r/2}
dx \   \frac{\D}{z^2}\sqrt{1-c^2z^8+ \sqrt{1-c^2z^8} {z^\prime}^2} \, ,
\ee
where $r$ is the distance between the quark and the anti-quark in the $x$ direction
and $\D$ is given by
\be
\D = \ls \frac{1 + c z^4}{1 - c z^4} \rs^{\sqrt{3/8}} .
\ee
Then the conserved Hamiltonian of this system can be obtained as
\be \label{ham1}
H = - \frac{\D}{z^2} \frac{1-c^2z^8}{\sqrt{1-c^2z^8+\sqrt{1-c^2z^8}{z^\prime}^2}} .
\ee
If there exists a point $z_0$ where $\left. \frac{\pa z}{\pa x} \right|_{z=z_0}$ becomes zero,
the Hamiltonian at this point becomes
\be \la{eq2}
H = - \frac{\D_0}{z_0^2} \sqrt{1-c^2 z_0^8} ,
\ee
where $\D_0$ is the value of $\D$ at $z=z_0$. Then the conservation law leads us to find
the relation between $r$ and $z_0$ from \eq{ham1} and \eq{eq2} as
\be \la{distanceQQ}
r=2  \int_0^{z_0} dz  \ \frac{z^2}{z_0^2} \frac{\sqrt{1-c^2 z_0^8}}{(1-c^2 z^8)^{1/4}}
\frac{1}{ \sqrt{\frac{\D^2}{\D_0^2}(1 - c^2 z^8)- \frac{z^4}{z_0^4}(1 - c^2 z_0^8)}} .
\ee
In addition, the quarkonium free energy proportional to the string on-shell action
becomes
\be \la{hmenergy}
E_{q} = 2
\int_0^{z_0} dz \ \frac{\D^2}{z^2 \D_0}
\frac{(1-c^2 z^8)^{3/4}}{ \sqrt{\frac{\D^2}{\D_0^2} (1 - c^2 z^8)
- \frac{z^4}{z_0^4} (1 - c^2 z_0^8)}} .
\ee
Since we consider the static string configuration in which the kinetic energy of quarks vanishes,
the above free energy can be interpreted as a sum of the quark masses and the interaction energy between
two heavy quarks. Notice that the quarkonium free energy in \eq{hmenergy} diverges
at the boundary $z = 0$. This is due to the infinite mass of a heavy quark in the hard wall model.
To understand this, we consider a straight string configuration. Under
the following ansatz
\be
\ta=t, \quad \sigma=z  \quad {\rm and} ~~ x_1 = {\rm const} ,
\ee
the free energy of two straight strings extended from $0$ to $z_{IR}$ in $z$-direction, which
corresponds to a mass of two heavy quarks in QCD side, is reduced to
\be
E_{m} = 2\int_0^{z_{IR}} dz \frac{\D (1-c^2 z^8)^{1/4} }{z^2} .
\ee
From this, we can see that the heavy quark in this model has an infinite mass.
Therefore to obtain a finite interaction energy we should calculate the free energy difference of two
string configurations, which we call effective potential
\be
V_{F1} = E_{q} - E_{m} .
\ee

If the above effective potential is bigger than the mass of two light quarks,
the quarkonium should be broken to two heavy-light mesons. Furthermore, since
the mass of a light quark is very small we can ignore the effect of the light quark mass.
Then, $V_{F1} < 0$ (or $V_{F1} > 0$) implies that
the quarkonium is more stable (or unstable) than two heavy-light meson.
So the interquark distance $r$ at $V_{F1} = 0$ can be interpreted as a
dissociation length.
Figure 1 (a) shows the dissociation length of quarkonium depending
on the gluon condensation.

%%%%%%%%%%%%%%%%%%%%%%%%%%%%%%%%%%%%%%%%%%%%%%%%%%%%%%%%%%%%%%%%%%%%%%%%%%%
\begin{figure}
\begin{center}
\vspace{0cm}
\hspace{-1.cm}
{ \includegraphics[angle=0,width=0.5\textwidth]{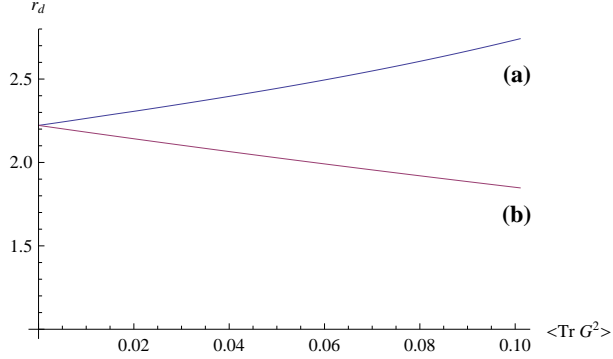}}
\caption{\small Dissociation length $r_d$ in the unit of GeV$^{-1}$, depending on the gluon condensation in the case of
(a) quarkonium and (b) the bound state of monopole-anti monopole in the hard wall approach.}
\label{number}
\end{center}
\end{figure}
%%%%%%%%%%%%%%%%%%%%%%%%%%%%%%%%%%%%%%%%%%%%%%%%%%%%%%%%%%%%%%%%%%%%%%%%%%%%

For describing the bound state of monopole and anti monopole, we consider a D1-string
instead of a fundamental string, whose action is given by \ct{Minahan:1998xb,Constable:1999ch}
\bea
S_{D1} &=& \int d^2 \s \ e^{- \ph} \sqrt{\det \ G^{MN} \pa_{\a} X^M
\pa_{\b} X^N} \nn
&=& \int_{-T/2}^{T/2} dt \int_{-r/2}^{r/2}
dx \frac{1}{z^2 \D}  \sqrt{1-c^2z^8+ \sqrt{1-c^2 z^8} {z^\prime}^2}  ,
\eea
where \eq{simple} is used. Following the same procedure used in the fundamental string case,
the interquark
distance $r$ and the free energy difference $E_{D1}$ between the monopole and the anti-monopole
can be written as functions of $z_0$ as follows:
\be    \la{distanceMM}
r = 2  \int_0^{z_0} dz  \ \frac{z^2}{z_0^2} \frac{\sqrt{1-c^2 z_0^8}}{(1-c^2 z^8)^{1/4}}
\frac{1}{ \sqrt{\frac{\D_0^2}{\D^2}(1 - c^2 z^8)- \frac{z^4}{z_0^4}(1 - c^2 z_0^8)}} ,
\ee
\bea
V_{D1} &=& 2
\int_0^{z_0} dz \ \frac{\D_0}{z^2 \D^2}
\frac{(1-c^2 z^8)^{3/4}}{ \sqrt{\frac{\D_0^2}{\D^2} (1 - c^2 z^8)- \frac{z^4}{z_0^4} (1 - c^2 z_0^8)}} \nn
&& - 2\int_0^{z_{IR}} dz \frac{(1-c^2 z^8)^{1/4} }{z^2 \D} .
\eea
Then the dissociation length of the monopole-anti monopole bound state depending on
the gluon condensation can be found numerically, see Figure 1 (b).
We find that the dissociation length of the quarkonium becomes large as the gluon
condensation increases, which implies that it is difficult to dissociate the quarkonium
due to the gluon condensation.
On the contrary, the monopole-anti monopole bound state can be
dissociated more easily as the gluon condensation increases.

\section{Braneless approach}

As mentioned previously, we can take into account another approach to study hadron physics in the confining phase, so called braneless
approach. In this case, since we use the intrinsic singular point $z_c$ as an IR cutoff, we do not need to introduce an artificial IR bound by hand.
Taking this to our advantage, we will now study the meson spectra in the braneless approach.

Note that all equations of motion for meson spectra are the same as the ones
in the hard wall case,
and the only difference is that the IR boundary position in the braneless approach
is given by $z=z_c$ instead of $z=z_{IR}$ .
Therefore, we will summarize the results for the meson spectra in the braneless
approach omitting the calculational details.

From \eq{eq:vector}, we see that the vector meson mass depends on the position of the singularity
only. So we can find the position of $z_c$, where the mass of the first excited mode $m_\r$ is assigned the  experimentally known value 776 MeV.
By choosing the boundary conditions, $V_i^{(1)} = 0 $ at $z=0$ and
$\pa_z V_i^{(1)} = 0$ at $z=z_c$,
 the position of the singular point is fixed as $z_c=1/325$MeV. Inserting this value into
\eq{gluoncond}, the value of the gluon condensation for $N_c=3$ is obtained as
\be
\bra \Tr G^2 \ket \approx 0.139 [{\rm GeV}^4].
\ee
It is larger than the the commonly expected value, $0.012 [{\rm GeV}^4]$
\ct{Miller:2006hr}. Moreover, as it is known  in the lattice QCD that  the gluon condensation in the confining phase
decreases as the temperature increases, it is interesting to investigate the meson masses varying gluon condensation.
In  Figure 2 we show the first and second excited vector meson masses.
Unlike the case of the hard wall
approach, the vector meson masses strongly depend on the gluon condensation. Moreover, the
vector meson masses increase as the gluon condensation increases, which is opposite
to the hard wall case.

%%%%%%%%%%%%%%%%%%%%%%%%%%%%%%%%%%%%%%%%%%%%%%%%%%%%%%%%%%%%%%%%%%%%%%%%%%%
\begin{figure}
\begin{center}
\vspace{2cm}
%\hspace{-1.cm}
\subfigure{ \includegraphics[angle=0,width=0.6\textwidth]{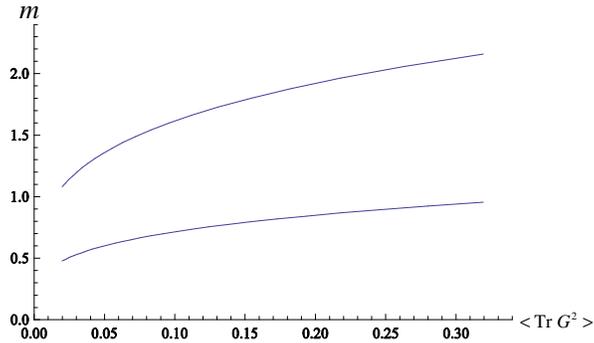}}
%\hspace{0.5cm}
%\subfigure[]{ \includegraphics[angle=0,width=0.45\textwidth]{decayvector.eps}}
\vspace{-4cm} \\
\caption{\small $\rho$-meson masses in GeV unit
of the first and second modes, $V_i^{(1)}$ and $V_i^{(2)}$, depending on the gluon condensation
in the braneless approach.}
\label{number}
\end{center}
\end{figure}
%%%%%%%%%%%%%%%%%%%%%%%%%%%%%%%%%%%%%%%%%%%%%%%%%%%%%%%%%%%%%%%%%%%%%%%%%%%%

The scalar field $v$ in the braneless approach has a log
divergence
at $z=z_c$ as shown in \eq{divergence}. Fortunately, the masses for the axial vector and pseudoscalar meson in \eq{eqs}
depend
only on $\G(z)$, the combination of the above logarithmic divergence and the metric component.
With the help
of the metric component, which vanishes at $z=z_c$, $\G(z)$ also becomes zero. As a result,
in all equations for the axial vector and pseudoscalar meson the divergent terms do not appear.
Imposing the boundary conditions, $\varphi = 0$, $\pi = 0$ at $z=0$
and $\pa_z \varphi = 0$ at $z=z_c$,
we find the masses of the axial vector and pseudoscalar meson
depending on the gluon condensation as drawn in Figure 3. For numerically exact values,
see Table 2.

%%%%%%%%%%%%%%%%%%%%%%%%%%%%%%%%%%%%%%%%%%%%%%%%%%
\begin{center}
\begin{tabular}{|c||c|c|c|c|c|c|}
\hline
 $z_c $ {\scriptsize (1/GeV)} & $ \langle {\rm Tr} G^2\rangle$  {\scriptsize (GeV$^4$)} & $ m_\rho$  {\scriptsize (GeV) } & $m_A$  {\scriptsize (GeV)} & $m_\pi$  {\scriptsize (GeV)}\\
\hline \hline
 1/0.200 &  0.020 & 0.4780 & 1.4081 & 0.13796 \\
\hline
 1/0.250 & 0.049 & 0.5975 & 1.4057 & 0.13808 \\
\hline
1/0.325&  0.139 & 0.7768 & 1.3574 & 0.14020 \\
\hline
 1/0.378 & 0.253 & 0.9035 & 1.2880 & 0.14743 \\
\hline
 1/0.400 & 0.319 & 0.9561 & 1.2715 & 0.15302\\
\hline
\end{tabular} \\
\vspace{0.3cm}
\noindent Table 2 : Meson masses depending on the gluon condensation in the braneless approach. \\
\end{center}

One can see from the Table 2 and the figures that the gluon condensation effect on the meson masses
in the braneless approach is not small unlike the cases of the hard wall approach. Furthermore,
the masses of the vector and axial vector meson in the braneless approach qualitatively behave
opposite to those in the hard wall approach.

%%%%%%%%%%%%%%%%%%%%%%%%%%%%%%%%%%%%%%%%%%%%%%%%%%%%%%%%%%%%%%%%%%%%%%%%%%%
\begin{figure}
\begin{center}
\vspace{0cm}
\hspace{-1.cm}
\subfigure[]{ \includegraphics[angle=0,width=0.45\textwidth]{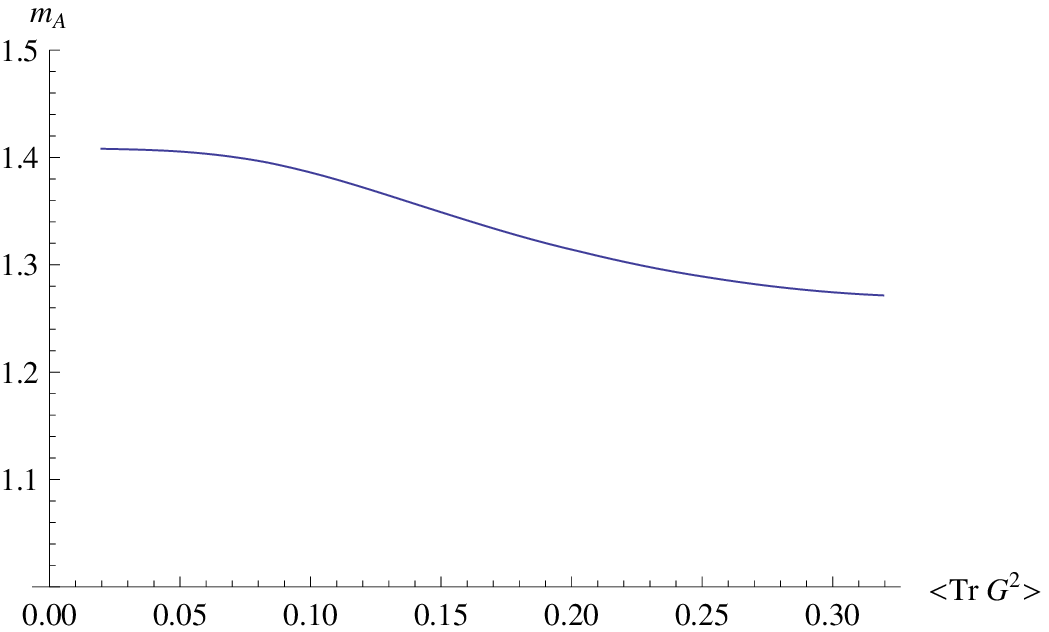}}
\hspace{0.5cm}
\subfigure[]{ \includegraphics[angle=0,width=0.45\textwidth]{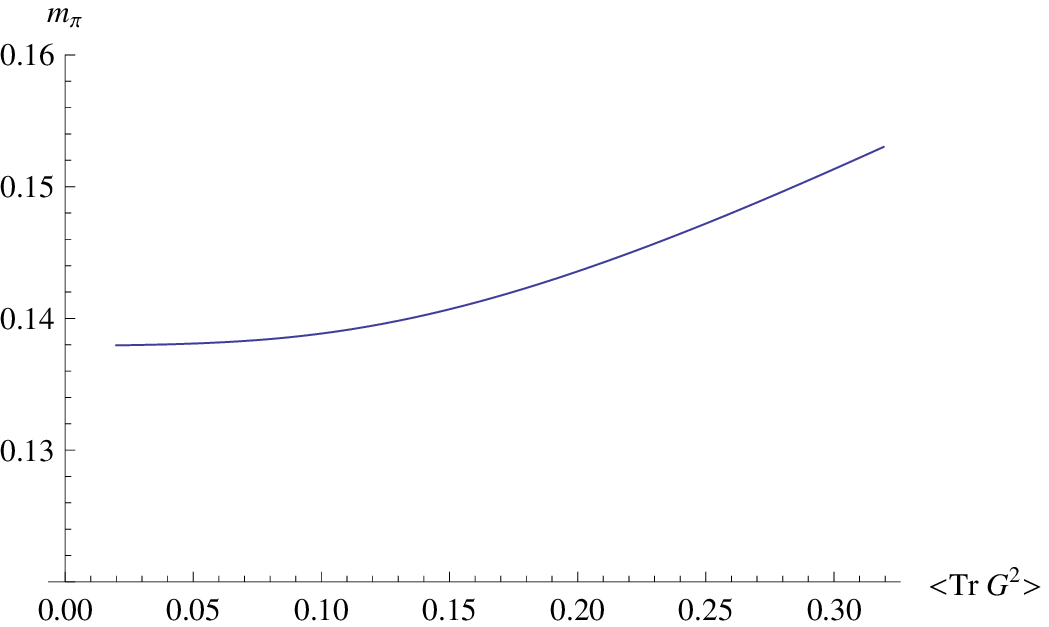}}
\vspace{0cm} \\
\caption{\small (a) axial vector meson mass and (b) pion mass in the unit of GeV depending on the
gluon condensate in the braneless approach.}
\label{number}
\end{center}
\end{figure}
%%%%%%%%%%%%%%%%%%%%%%%%%%%%%%%%%%%%%%%%%%%%%%%%%%%%%%%%%%%%%%%%%%%%%%%%%%%%

In the case of the dissociation of
the quarkonium and the monopole-anti monopole bound state, the distance between the quark
and anti-quark or the
monopole and anti-monopole is described by the same integral equations, \eq{distanceQQ} or
\eq{distanceMM}, respectively.
The free energy differences for quarkonium and monopole-anti monopole bound state are slightly changed into
\bea    \la{energy:braneless}
V_{F1} &=& 2
\int_0^{z_0} dz \ \frac{\D^2}{z^2 \D_0}
\frac{(1-c^2 z^8)^{3/4}}{ \sqrt{\frac{\D^2}{\D_0^2} (1 - c^2 z^8)- \frac{z^4}{z_0^4} (1 - c^2 z_0^8)}} \nn
&& - 2\int_0^{z_{c}} dz \frac{\D (1-c^2 z^8)^{1/4} }{z^2} , \nn
V_{D1} &=& 2
\int_0^{z_0} dz \ \frac{\D_0}{z^2 \D^2}
\frac{(1-c^2 z^8)^{3/4}}{ \sqrt{\frac{\D_0^2}{\D^2} (1 - c^2 z^8)- \frac{z^4}{z_0^4} (1 - c^2 z_0^8)}} \nn
&& - 2\int_0^{z_{c}} dz \frac{(1-c^2 z^8)^{1/4} }{z^2 \D} .
\eea

From \eq{distanceQQ}, \eq{distanceMM} and \eq{energy:braneless}, the dissociation lengths of the
quarkonium and monopole-anti monopole bound state are represented in Figure 5 (a) and (b).
Unlike the hard wall approach case, the dissociation of the quarkonium and
monopole-anti monopole bound state decreases as the gluon condensation increases.

\section{Summary and Discussion}

%%%%%%%%%%%%%%%%%%%%%%%%%%%%%%%%%%%%%%%%%%%%%%%%%%%%%%%%%
\begin{figure}
\begin{center}
\vspace{0cm}
\hspace{-1.cm}
{ \includegraphics[angle=0,width=0.5\textwidth]{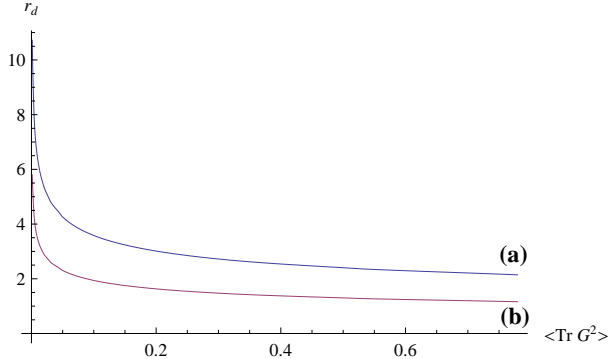}}
\hspace{0.5cm}
\caption{\small Dissociation length, $r_d$ in the unit of GeV$^{-1}$, depending on the gluon condensation in the case of
(a) quarkonium and (b) the bound state of monopole-anti monopole in the braneless approach.}
\label{number}
\end{center}
\end{figure}
%%%%%%%%%%%%%%%%%%%%%%%%%%%%%%%%%%%%%%%%%%%%%%%%%%%%%%%%%
We have studied meson spectra on the gluon condensate background using two alternative
approaches, termed  ``hard wall'' and  ``braneless'', in the holographic QCD model. The backreaction of the
scalar field corresponding to the gluon condensation usually causes a divergence, which
gives a naked singularity in the bulk geometry.

In the hard wall model, we introduced an IR cutoff by hand in order to describe the confining phase and assumed that an IR cutoff exists ``in front''
of the singularity. In this case,  because the IR cutoff screens the singularity, we do not need to worry about the existence of the
singularity. If the IR cutoff lies behind the singularity, the results become those of the braneless approach. We found that the quantitative change of meson masses does not strongly
depend on the gluon condensate in the hard wall approach. Qualitatively, the vector meson mass
decreases slightly as the gluon condensation increases. On the contrary,
masses of the axial vector meson and pion increase very
slowly. Furthermore, we also found that the dissociation length of the quarkonium and
monopole anti-monopole bound state
becomes respectively larger and shorter, as the gluon condensation increases.

In the braneless approach, where the singularity was identified with the IR cutoff,
we have shown that the equations for meson spectra are well defined
in spite of the singularity. Due to the merit of the braneless approach, in which we do not
need to introduce an IR cutoff by hand, we also investigated the meson spectra.
In this approach, we found that meson masses similar to those in the EKSS model
\cite{Erlich:2005qh} can be obtained
for $\bra \Tr G^2 \ket =0.139 {\rm [GeV]}^4$, which is bigger than the commonly expected value
of the gluon condensation, $0.012 {\rm [GeV]}^4$. Moreover, we found that the meson spectra
significantly depend on the gluon condensate and that
masses of the vector meson and pion increase as the gluon condensation becomes large, while
the axial vector meson mass decreases. For the dissociation of the
quarkonium and monopole anti-monopole bound state it was shown that the dissociation length
of both increases as the gluon condensate increases. This implies that
it gets more difficult to dissociate a quarkonium and monopole anti-monopole bound states for
a large value of the gluon condensate.

The two alternative approaches considered here give totally different results for
 the meson masses.
As we expect that the gluon condensate may not affect
the meson spectra strongly,
the hard wall model might be more appropriate to describe the gluon condensate
than the braneless approach.
To clarify this further, it would be interesting to compare our results with observations in future experiments. Our work might be helpful for understanding the gluon condensation effect
in real QCD.

\vspace{1cm}

{\bf Acknowledgement}

This work was supported by the National Research Foundation of Korea(NRF) grant funded by the
Korea government(MEST) through the Center for Quantum Spacetime(CQUeST) of Sogang
University with grant number 2005-0049409."
\vspace{1cm}

%%%%%%%%%%%%%%%%%%%%%%%%%%%%%%%%%%%%%%%%%%%%%%%%%%%%%%%%%%%%%%%%%%%%%%%%%%%%%%
%                                                                            %
%   The Bibliography                                                         %
%                                                                            %
%%%%%%%%%%%%%%%%%%%%%%%%%%%%%%%%%%%%%%%%%%%%%%%%%%%%%%%%%%%%%%%%%%%%%%%%%%%%%%

\end{document}